# A Model for Enhancing Human Behaviour with Security Questions: A Theoretical Perspective

**Nicholas Micallef**
Australian Centre for Cyber Security
School of Engineering and Information Technology
University of New South Wales
Canberra, Australia
Email: n.micallef@adfa.edu.au

**Nalin Asanka Gamagedara Arachchilage**
Australian Centre for Cyber Security
School of Engineering and Information Technology
University of New South Wales
Canberra, Australia
Email: nalin.asanka@adfa.edu.au

## Abstract

Security questions are one of the mechanisms used to recover passwords. Strong answers to security questions (i.e. high entropy) are hard for attackers to guess or obtain using social engineering techniques (e.g. monitoring of social networking profiles), but at the same time are difficult to remember. Instead, weak answers to security questions (i.e. low entropy) are easy to remember, which makes them more vulnerable to cyber-attacks. Convenience leads users to use the same answers to security questions on multiple accounts, which exposes these accounts to numerous cyber-threats. Hence, current security questions implementations rarely achieve the required security and memorability requirements. This research study is the first step in the development of a model which investigates the determinants that influence users' behavioural intentions through motivation to select strong and memorable answers to security questions. This research also provides design recommendations for novel security questions mechanisms.

# 1   Introduction

Internet users are increasingly dealing with more online accounts (statistics show that 92% of Australians use the internet (Poushter 2016)), for personal emails, social networks, e-commerce, banks etc. Hence, internet users are finding it more challenging to remember the passwords of all of their online accounts (Florencio and Herley 2007; Stavova et al. 2016). Recent research found that password managers have not been widely adopted (Alkaldi and Renaud 2016). Thus, resetting passwords is increasingly becoming a much more frequent task (Florencio and Herley 2007; Stavova et al. 2016).

Various types of fall-back authentication mechanisms have been studied to address this problem with password recovery mechanisms (Stavova et al. 2016). The most popular being email-based password recovery, text-based password recovery and security questions (Schechter and Reeder 2009). Although both email-based and text-based password recovery have recently been adopted by major companies (e.g. Google and Facebook), they still have major limitations (e.g. security vulnerabilities and lack of mobility) (Stavova et al. 2016). For instance, with email-based password recovery users might require to pay a higher cost for data roaming if they need to recover their passwords, when they are abroad. Also, for text-based password recovery, users might not carry their device with them when on vacation. In this situation, it would be impossible for users to recover their forgotten password.

From a security questions perspective, weak answers to security questions (i.e. low entropy) are easy to remember (Zviran and Haga 1990), which makes them more vulnerable to the art of human hacking (i.e. guessing attacks, dictionary attacks, observational attacks and shoulder surfing attacks) (Bonneau et al. 2010; Denning et al. 2011). Instead, strong answers to security questions (i.e. high entropy) are less vulnerable to cyber-attacks (Shay et al. 2012), but at the same time are difficult to remember (Micallef and Just 2011). Therefore, convenience leads users to use the same answers to security questions on multiple accounts (Honan 2012), which exposes these accounts to numerous cyber-threats.

One could argue, that the main problem with security questions is that they are poorly designed, since more effort has been invested in improving the security and usability of passwords (Bonneau et al. 2012) (e.g. password meters (Komanduri et al. 2011), password rules). While in comparison, research on improving the security and memorability of security questions is still quite limited (e.g. there are no websites that provide security questions answer meters) (Senarath et al. 2016). Research on designing secure and usable security questions might still be limited due to the fact that passwords are considered to be more important, since passwords are used as the main mechanism to login into an online account (Bonneau and Preibusch 2010). Alternatively, security questions are mostly used for password recovery (Just and Aspinall 2010). However, if an online account could be breached through weak answers to security questions than it does not matter how strong and usable a password login mechanism is because the attacker could still get control of the account and consequently reset the password to use the account for malicious purposes. For instance, in Sarah Palin's 2008 email account hack, the attacker merely used social engineering techniques to reset her password using her birth-date, ZIP code and where she met her spouse (Bridis 2008). In 2014, Apple released a statement which stated that certain iCloud accounts were compromised by a targeted attack on user names, passwords, and security questions (Albanesius 2014). Therefore, equal importance (as for designing usable and secure passwords) should be given to designing secure and usable security questions. Therefore, since other fall-back authentication mechanisms (e.g. text-based password recovery) have their own limitations (e.g. security vulnerabilities and lack of mobility) (Stavova et al. 2016), designing both secure and usable security questions needs to be investigated in a more pragmatic way.

To design both secure and usable security questions mechanisms we need to better understand the human behaviour. Therefore, in this research, we focus on investigating which are the determinants that influence users' behavioural intentions through motivation to select strong and memorable answers to security questions. Hence, in our work, we use previous research on Protection Motivation Theory (PMT) (Rogers 1975; Maddux and Rogers 1983) and Technology Threat Avoidance Theory (TTAT) (Liang and Xue 2010; Arachchilage and Love 2013, 2014) to define a model that aims to investigate users' behavioural intentions when selecting answers to security questions to protect themselves from cyber-attacks. We also use previous research on human memory (Atkinson and Shiffrin 1968; Anderson and Bower 1972) to enhance the proposed model, so that it takes into consideration memorability traits that affect users' behaviour intentions when selecting answers to security questions. Finally, we discuss how the proposed research model could affect the design of a novel security questions mechanism.

The remainder of this paper is organised as follows. The 'Theoretical Background' section illustrates the basics of PMT (Rogers 1975) and TTAT (Liang and Xue 2010). The 'Research Model' section, presents the proposed research model and motivations behind the design of the model and related hypotheses. In Section 4, we identify the implications that the proposed model would have on the design of a novel security questions mechanisms that aims to enhance human behavioural intentions through motivation when selecting answers to security questions. Finally, conclusion and future research regarding the proposed research model are presented in Section 5.

## 2  Theoretical Background

The aim of this research is to design a model that investigates the determinants that influence human behavioural intentions through motivation to select strong and memorable answers to security questions. In this research we construct the research model by exploring the two main aspects that could motivate users to select strong and memorable answers to security questions: (1) behavioural traits that influence security and (2) behavioural traits that influence usability.

With respect to the behavioural traits that influence security, previous research defined the Protection Motivation Theory (PMT) (Rogers 1975; Maddux and Rogers 1983). PMT interprets why and how people decide to take protective behaviours. Current research in security behaviour adopted the PMT model by creating the Technology Threat Avoidance Theory (TTAT), through the use of health behavioural models and risk analysis research (Liang and Xue 2010). According to Liang and Xue (2010) (and other research in the field (Arachchilage and Love 2013; Arachchilage et al. 2016; Tsai et al. 2016)), TTAT implements PMT's framework by detecting the main factors that lead to technology threat avoidance behaviour. This is achieved by exploring the relationship between threat (in this case, the threat of hacking online accounts due to weak answers to security questions) and coping (in this case, measures taken to avoid hacking of online accounts which use security questions) appraisals and their role in motivating the use of protective behaviours, as instigated by risk tolerance and social influences. Previous research (Liang and Xue 2010; Arachchilage and Love 2013; Arachchilage et al. 2016; Tsai et al. 2016) found that threat appraisals (perceived susceptibility and severity) and coping appraisals (safeguard effectiveness, safeguard cost, and self-efficacy) were significant predictors of computer threat avoidance behaviours. Since the aim of our research is to design a model that investigates the determinants that could influence users' motivation to improve their behavioural intentions when selecting strong answers to security questions, in order to protect their accounts from cyber-attacks, we also use TTAT as a means of extending PMT.

Previous research found that internet users are more inclined to get discouraged in taking measures to cope with a threat, which could arise from using weak answers to security questions because they find the cost of remembering strong answers to security questions to be prohibitive (Schechter and Reeder 2009; Schechter et al. 2009). Hence, in our research we also investigate behavioural traits that influence usability (by focusing specifically on memorability, since for security questions memorability is the main determinant of usability (Just and Aspinall 2009; Bonneau et al. 2015)), as a separate entity. The aim is to further understand how memorability affects the avoidance motivation and consequently the improved behavioural intentions of using strong and memorable answers to security questions. Since previous work found that entertaining ads tend to be more memorable (Kellaris and Cline 2007) than non-entertaining ads, our research also investigates whether entertainment would have the same effect on remembering answers to security questions. Entertainment is important because recent research (Micallef and Arachchilage 2017a, 2017b) has started evaluating the use of games with the aim of enhancing the memorability of answers to security questions. In the proposed research model, we also add social influences because Johnston and Warkentin (2010) found that social influences are important to understand how others behave. No other research has investigated in detail the behavioural traits that influence both security and usability to motivate users to select stronger and memorable answers to security questions.

## 3  Research Model

The proposed research model (see Figure 1) aims to enhance human behaviour with security questions by exploring the two aspects (i.e. security and usability) that could influence users' motivation to improve their behavioural intentions to select strong and memorable answers to security questions. The part of the model which refers to the behavioural traits which influence security was derived from Liang and Xue (2009) and also adopted and evaluated by plenty of other research in the security behaviour field (Liang and Xue 2010; Arachchilage and Love 2013, 2014; Tsai et al. 2016). In line with TTAT, we propose that internet users' cyber-threat avoidance behaviour is determined by avoidance motivation. Avoidance motivation is affected by perceived threat, safeguard effectiveness, safeguard cost and self-efficacy. However, since for security questions memorability is the main determinant of usability (Just and Aspinall 2009; Bonneau et al. 2015), we propose that avoidance motivation is also affected by the memorability of answers to security questions (see behavioural traits which influence usability in Figure 1). The following sections explain in more detail the reasoning behind the choices that were taken when designing the proposed research model (see Figure 1) and corresponding hypotheses.

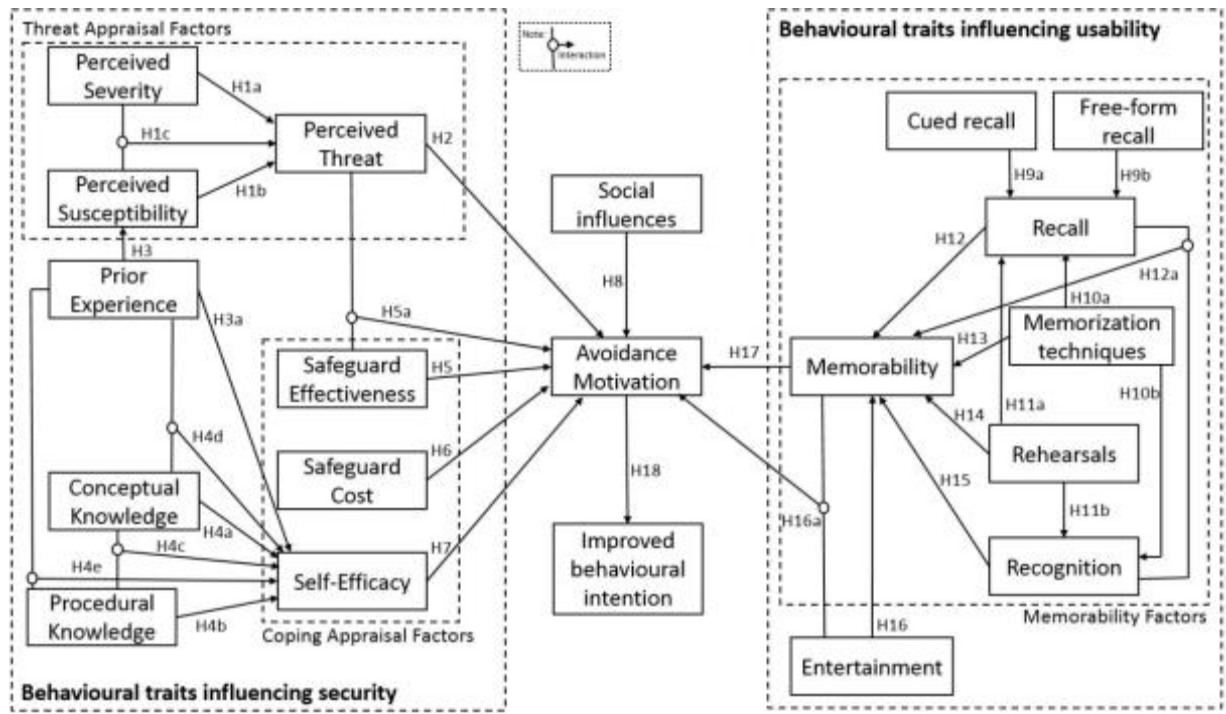

*Figure 1: Proposed research model (enhanced from TTAT (Liang and Xue 2010))*

## 3.1 Behavioural Traits Influencing Security

A perceived threat is defined as the extent to which a person perceives a situation to be dangerous or harmful (Liang and Xue 2010). Perception of threat is developed by examining the personal computing environments to detect potential dangers (Liang and Xue 2010). According to previous research (Liang and Xue 2010; Arachchilage and Love 2013), perceived susceptibility and perceived severity influence the threat perception. Perceived severity is the magnitude to which an internet user perceives that negative consequences caused by malicious behaviour will be severe (Liang and Xue 2010). Perceived susceptibility is the internet users' personal opinion on whether malicious behaviour will affect him or not (Liang and Xue 2009). Most security behaviour research (Liang and Xue 2010; Arachchilage and Love 2013; Tsai et al. 2016; Boss et al. 2015) agree that susceptibility and severity seem to determine the security threats that internet users face continuously. Therefore, in our research model we propose that even for threats that arise from cyber-attacks on accounts that use security questions, both perceived susceptibility and severity contribute to threat perception (**H1a**, **H1b**).

Previous research (Liang and Xue 2010; Arachchilage and Love 2013; Arachchilage et al. 2016) found that perceived susceptibility and severity have an interaction effect on threat perception. An interaction effect is a multiplication of two variables (i.e. perceived susceptibility and severity). Hence, perceived severity moderates the relationship between perceived susceptibility and perceived threat, and the other way round. When the perceived susceptibility of a malicious behaviour is high, internet users are more sensitive to changes in its severity level. Therefore, in our research model we propose that even for threats that could arise from cyber-attacks on online accounts that use security questions, perceived susceptibility moderates the perceived severity of the threat perception, and the other way round (**H1c**).

When internet users perceive a threat, they are motivated to avoid it (Liang and Xue 2010). Avoidance motivation is defined as the extent to which internet users are motivated to avoid cyber-threats by taking safeguarding measures (e.g. using stronger answers to security questions). As the threat perception increases, internet users get more motivated to avoid the danger. The vulnerabilities of security questions (Just and Aspinall 2010) towards numerous cyber-threats have increased. This is leading to severe consequences, such as embarrassment, reputation and monetary loss (Micallef and Just 2011). Hence, based on empirical evidence from other research conducted on security avoidance behaviour (Liang and Xue 2010; Arachchilage and Love 2013; Arachchilage et al. 2016; Tsai et al. 2016), in our research model we propose that the perceived threat of cyber-attacks on online accounts which use security questions affects avoidance motivation (**H2**).

Prior experience (in this case, with cyber-attacks on online accounts) was not included in the original PMT model (Rogers 1975) or the adopted TTAT model (Liang and Xue 2010). Although, eventually it was suggested to add prior experience into the PMT model (Maddux and Rogers 1983), few studies have actually added this variable (Mwagwabi et al. 2014; Tsai et al. 2016). Since Mwagwabi et al. (2014) found that prior experience (in their case, hacking exposure) does affect the perceived vulnerability when complying with password guidelines, we incorporate prior experience as part of the threat appraisals of our research model (**H3**). Also, since previous research (Zhao et al. 2005) found that prior experience (i.e. previous entrepreneurial experience affected students' intentions to become entrepreneurs) does affect self-efficacy, in our research model we propose that prior experience does affect self-efficacy, to enhance threat avoidance behaviour when selecting answers to security questions (**H3a**).

Conceptual knowledge is the *"know what"* of knowledge, while procedural knowledge is the *"know how"* of knowledge (Plant 1994). McCormick (1997) argued that learning procedural and conceptual knowledge of technological activities can influence knowledge. Previous research (McCormick 1994; Arachchilage and Love 2014) argue that conceptual and procedural knowledge should be considered separately, however, there is a relationship between them. Other research (Hsu et al. 2007; Hu 2010) found that knowledge affects self-efficacy. For example, when users are knowledgeable of the threats that could arise from using weak answers to security questions, they are more confident to take relevant actions to avoid potential cyber-attacks. Thus, in our research model we propose that both procedural and conceptual knowledge as well as their interaction effect positively affects self-efficacy, which contributes to enhance threat avoidance behaviour when selecting answers to security questions (**H4a, H4b, H4c**). According to Bonner and Walker (1994) experience is one of the factors that determines the acquisition of knowledge. For example, prior experience of having an online account hacked, leads a user to gather knowledge about what lead to the hacking incident and how to prevent another occurrence. Hence, in our research model we propose that both procedural and conceptual knowledge have a positive interaction effect with prior experience towards self-efficacy, which contributes to enhance threat avoidance behaviour when selecting answers to security questions (**H4d, H4e**).

When internet users perceive a threat, they start considering coping measures to potentially safeguard themselves from that threat. Based on previous research (Liang and Xue 2010; Arachchilage and Love 2013; Boss et al. 2015; Arachchilage et al. 2016; Tsai et al. 2016), internet users' avoidance motivation of a perceived threat is determined by three variables: safeguard effectiveness, safeguard cost and self-efficacy. Safeguard effectiveness is defined as the individual's assessment of a safeguarding measure regarding how effectively it can be applied to avoid malicious cyber-threats (Liang and Xue 2010; Arachchilage and Love 2013). For example, the individual assessment regarding how effectively strong answers to security questions could be used to prevent cyber-attacks on online accounts. Safeguard cost is a payback for safeguard effectiveness. This refers to the physical and cognitive efforts such as time, money, inconvenience and comprehension required using safeguarding measures (Liang and Xue 2009). Self-efficacy is defined as individuals' confidence in taking the safeguarding measure. This is an important determinant of avoidance motivation. Previous research has found that internet users are more motivated to perform IT security related behaviours as the level of their self-efficacy increases (Woon et al. 2005; Ng et al. 2009; Liang and Xue 2010; Arachchilage and Love 2013; Arachchilage et al. 2016; Tsai et al. 2016). Therefore, in our research model we propose that even for threats arising from attacks on online accounts which use security questions, coping mechanisms consist of: safeguard effectiveness, safeguard cost and self-efficacy (**H5, H6, H7**).

Liang and Xue (2010) found that perceived threat can negatively moderate the relationship between safeguard effectiveness and avoidance motivation. Knowing that the safeguarding measure can effectively reduce the threat, an internet user will not be so eager to cope with it, although the user is fully aware of the threat. Therefore, in our research model we propose that as the level of safeguard effectiveness increases, users tend to feel less motivated to avoid the threat (**H5a**).

### 3.2　Behavioural Traits Influencing Memorability

To define the hypotheses related to memorability factors for our proposed research model, we examine previous work that studied in detail how the human memory functions. Atkinson and Shiffrin (1968) proposed a cognitive memory model, in which, new information is transferred to short-term memory through sensory organs. The short-term memory holds this new information as mental representations of selected parts of the information. This information is only passed from short-term memory to long-term memory when it can be encoded through cue-association (Atkinson and Shiffrin 1968) (e.g. seeing a cat crossing the street reminds us of our first cat). This encoding through cue-association helps people to remember and retrieve the stored information over an extended period of time. These encodings are strengthened through constant rehearsals. Therefore, in our research model we propose that both encoding (memorization) and rehearsals improve the memorability of answers to security questions, even for recall and recognition tasks (**H10a, H910b, H11a, H11b, H13, H14**).

Recall and recognition are the processes that are used to retrieve information from memory (Stobert and Biddle 2013). Recognition is the process of remembering contextual information when a focus is provided, while recall is the process of remembering a specific focus when context is provided (Hollingworth 1913). Recall can either happen through cued-recall (e.g. a picture of a school would remind us of our first school), where a cue aids the retrieval of information, or free recall (e.g. recalling the password of an email account), where no aid is provided (Stobert and Biddle 2013). Therefore, in our research model we propose that cued recall has a positive effect on recalling answers to security questions while free-form recall has a negative effect (**H9a, H9b**).

Researchers argue that recognition is easier to perform when compared to recall (Haist et al. 1992). According to the generate-recognize theory (Anderson and Bower 1972), recall is a two phase process: Phase 1 (generate) - A list of possible words is formed by looking into long-term memory; Phase 2 (recognize) - The list of possible words is evaluated to determine if the word that is being looked for is within the list. According to this theory, recognition does not use the first phase which makes it easier and faster to perform. The theory also explains the benefits of cueing on memory retrieval. A cue helps both when generating a relevant candidate list, but also in recognizing the appropriate word from that list (Stobert and Biddle 2013). Therefore, in our research model we propose that recall has a negative effect on memorability, while recognition has a positive effect on memorability (**H12, H15**).

Based on previous memorability research one could argue that our proposed research model should only contain recognition and omit recall. However, using only recognition would decrease the level of security of answers to security questions because it would provide a limited answer space when compared to recall (Micallef and Arachchilage 2017a, 2017b). Therefore, since the aim of this research is not only to strengthen security, but also to strengthen memorability, we argue that the proposed research model should also study the interaction effect between recall and recognition to enhance memorability, but at the same time don't undermine security (**H12a**).

Previous research (Crawford et al. 2013) found that users are ready to trade-off security for convenience (i.e. usability) during smartphone authentication (Micallef et al. 2015). This trade-off between security and usability is also common in various other scenarios, such as using same answers to security questions for multiple accounts (Honan 2012). Hence, even for answers to security questions, usability (in terms of memorability) negatively affects the users' motivation to create strong answers to security questions. Therefore, in our research model we propose that memorability negatively affects the avoidance motivation of selecting strong answers to security questions (**H16**).

### 3.3 Entertainment

Previous work in the advertising field found that entertaining ads tend to be more memorable (Kellaris and Cline 2007) than ads which are not entertaining. This is further backed up by other research (Ho et al. 2009), which evaluated the use of tangible user interfaces (e.g. WII remote control), when playing a health education-based game. They found that an increase in fun and enjoyment leads to an increased memorability and consequently an improved learning experience. One could argue that an improved learning experience provides a high user satisfaction, which was empirically investigated by previous work on security behaviour (Arachchilage et al. 2016), that it could motivate users to change peoples' phishing threat avoidance behaviour. Therefore, in our research model we propose that entertainment positively affects remembering answers to security questions and that there is a positive interaction between entertainment and memorability with respect to avoidance motivation (**H16, H16a**).

### 3.4 Social Influences

Internet users are usually part of an organization, a group or a society. Thus, their behaviours could be influenced by other people. Liang and Xue (2009) found that computer users' threat avoidance behaviour could be influenced by their social environment (i.e. family, friends). This happens because when internet users do not know how to avoid a cyber-threat, they turn to their social connections to help them come up with safeguarding measures (Tu et al. 2014). Lee and Larsen (2009) examined in detail how social influences affects internet users' behaviour and found that IT executives are more likely to adopt anti-malware software in their organisations if their social connections (including business connections, customers, business partners) used the software. Therefore, in our research model social influences have a positive effect on avoidance motivation (**H8**).

### 3.5　Improved Behaviour

Consistent with previous TTAT research on security behaviour (Liang and Xue 2010; Arachchilage and Love 2013), in our research model we do not differentiate between avoidance motivation and improved behavioural intention. This is justified by the fact that behavioural intention is a strong predictor of actual behaviour (Ajzen 1991). This relationship has also been confirmed by technology adoption studies (Venkatesh et al. 2003; Pavlou and Fygenson 2006). Therefore, to be consistent with previous research, we argue that internet users which have a strong avoidance motivation are very likely to engage in an improved avoidance behaviour to protect their online accounts from cyber-attacks arising from weak answers to security questions (**H18**).

## 4　Designing a Novel Security Questions Mechanism: An Avatar Profile Case Study

A possible way to reduce the vulnerabilities of security questions towards the art of human hacking is to encourage users to use system-generated answers to security questions (Micallef and Just 2011). Previous research (Micallef and Just 2011) proposed a novel security questions mechanism, that uses an Avatar to represent system-generated data of a fictitious person, and then the Avatar's system-generated data is used to answer security questions (Micallef and Just 2011). Although this mechanism has not been extensively investigated, Micallef and Arachchilage (2017a, 2017b) argue that it has the potential to improve the trade-off between security and usability (in terms of memorability) due to the following reasons: (1) it could be tailored for a wide range of people; (2) guessing attacks could be minimized because the entropy and variety of answers could be defined/controlled by the system that generates them; (3) risks of having observational attacks would be minimal since the system-generated avatar information would not be publicly available; and (4) memorability could be improved by using a gamified approach to create and nurture a bond between users and their avatar profiles (in the form of system-generated data). In this research, we argue that the design of a novel security questions mechanism which uses avatar profiles (Micallef and Just 2011) could be further improved by using the following design recommendations that were extracted from the proposed research model (Figure 1).

Based on the behavioural traits which influence security (see left part of Figure 1), we suggest that a novel security questions mechanism which uses avatar profiles should have features that enable reminders of previous experiences (involving cyber-attacks of online accounts, which might have affected users themselves or close friends/acquaintances). This is due to the fact that according to our proposed research model prior experiences seem to affect the perceived susceptibility towards a threat (Mwagwabi et al. 2014), but also self-efficacy together with interaction effects with conceptual and procedural knowledge (Bonner and Walker 1994). This feature could be implemented either through asking questions about prior experiences (Briggs et al. 2016), during the avatar profile generation phase or through nudging users with periodic notifications (Almuhimedi et al. 2015; Felt et al. 2015), when logging into an account which uses the details of the selected avatar profile to recover passwords. Previous work conducted on fall-back authentication mechanisms (Micallef and Just 2011; Shay et al. 2012; Hang et al. 2015) have not leveraged the use of previous experiences to improve self-efficacy or increase the susceptibility of threats that could arise from using weak answers to security questions.

According to the behavioural traits that influence security (see left part of Figure 1), users' threat perception could also be increased by affecting their perceived susceptibility and severity (Liang and Xue 2010; Arachchilage and Love 2013; Tsai et al. 2016; Boss et al. 2015). Besides notifications or questions related to the users' prior experiences, an avatar security questions mechanism should also nudge users' perception towards cyber-threats launched on online accounts of victims that use weak answers to security questions (Briggs et al. 2016). Thus, it is suggested that notifications (i.e. messages) (Felt et al. 2015) should appear when logging into accounts, which use avatar profiles to recover passwords. These accounts should also show scenarios in which other accounts were hacked using weak answers to security questions. Furthermore, the interface should describe the severe consequences that victims suffered due to these attacks.

Research (Bonneau at al. 2015) suggests that sometimes users do not seem to realize that the answers to their security questions are not effective in safeguarding themselves from cyber-attacks (Bonneau at al. 2015). Therefore, the proposed research model implies that a novel security questions mechanism that uses avatar profiles should show security questions answers meters (as password meters (Komanduri et al. 2011)). This feature would help users understand the strength of the answers of the security questions that they selected (Senarath et al. 2016).

Our research model (see Figure 1) proposes that social influences affect one's behaviour (Tu et al. 2014). Thus, a novel security questions mechanism that uses avatar profiles should implement features that encourage social interactions. For instance, this mechanism could have features which allow users to use social media (e.g. Facebook, Twitter or LinkedIn) to notify their social acquaintances that they are using mechanisms that help them strengthen and memorize their answers to security questions, with the purpose of being less vulnerable to cyber-attacks. In the Mobile HCI field, previous work on improving the adherence to rehabilitation conditions has highlighted the importance of social connections on motivation (Micallef et al. 2016). However, despite other areas of research highlighted the importance of social connections to enhance motivation, previous work on fall-back authentication mechanisms (Denning et al. 2011; Juang et al. 2012; Bonneau and Schechter 2014) have not evaluated features which involve social interactions to influence threat avoidance motivation for security purposes. Thus, we suggest that the use of social interactions should be investigated in more detail in interfaces and systems (e.g. security questions) that try to educate people to change their security behaviour.

The behavioural traits that influence usability (see right part of Figure 1) of the proposed research model suggest that a novel security questions mechanism which uses avatar profiles should implement features, through the use of memorization techniques, with the aim of helping users store information in long-term memory (Atkinson and Shiffrin 1968). For example, after selecting the attributes from the avatar profile that users would use to answer security questions, memorization exercises could also be provided to help users memorize the information (Juang et al. 2012). Our research model also proposes the use of rehearsal techniques. This is due to the fact that rehearsals help strengthen the storage of information in long-term memory and the stronger the association with the information the more likely it is that users will remember the answers to their security questions (Micallef and Arachchilage 2017a, 2017b). This feature could be implemented by sending reminders (i.e. messages) to the users to have a look at their avatar profile, in order to refresh their memory of the attributes that they selected to answer their security questions. Based on previous memorability research (Atkinson and Shiffrin 1968; Haist et al. 1992), the proposed research model suggests that a novel security questions mechanism which uses avatar profiles should implement cued recall to minimize the instances in which free-form recall is required. For example, an image of the selected avatar could be shown at each login to facilitate cued-recall (Denning et al. 2011; Stobert and Biddle 2012; Castelluccia et al. 2017).

The design recommendations that were discussed in this section were particularly tailored for a novel security questions mechanism which uses avatar profiles (Micallef and Just 2011). However, we suggest that these recommendations should also be implemented in fall-back authentication mechanisms (i.e. security questions) and cyber security education (game-based learning approach for security) systems, which also aim to reduce the trade-off between security and usability. Based on the theoretical perspective that we used to define the proposed research model, we strongly believe that these design recommendations could potentially enhance users' behavioural intentions through motivation to select strong and memorable answers to security questions.

## 5  Conclusion and Future Research

The main contribution of this research is the definition of a model that investigates the determinants that could influence the behavioural intentions that motivate users to select strong and memorable answers to security questions. The proposed research model was constructed using previous research on PMT (Rogers 1975) and TTAT (Liang and Xue 2010) to investigate the behavioural traits that users use when selecting answers to security questions, to protect themselves from cyber-attacks. The proposed research model also uses previous research on human memory (Atkinson and Shiffrin 1968) to investigate the behavioural traits which affect memorability (the strongest determinant for usability (Just and Aspinall 2009)) when selecting answers to security questions. Finally, we discuss how the proposed research model could improve the design of novel security questions mechanisms that aim to enhance human behavioural intentions by motivating users to select strong and memorable answers to security questions. In future work we will empirically verify the validity of this research model.